\documentclass[a4paper]{jpconf}
\usepackage{graphicx}
\begin{document}
\title{Thermal pairing and giant dipole resonance in highly excited nuclei}

\author{Nguyen Dinh Dang}

\address{1) Theoretical Nuclear Physics Laboratory, Nishina Center for Accelerator-Based Science, RIKEN, 2-1 Hirosawa, Wako city, 351-0198 Saitama, Japan}
\address{2) Institute for Nuclear Science and Technology, Hanoi, Vietnam}

\ead{dang@riken.jp}

\begin{abstract}
Recent results are reported showing the effects of thermal pairing in highly excited nuclei. It is demonstrated that thermal pairing included in the phonon damping model (PDM) is responsible for the nearly constant width of the giant dipole resonance (GDR) at low temperature $T <$ 1 MeV.
It is also shown that the enhancement observed in the recent experimentally extracted nuclear level densities in $^{104}$Pd at low excitation energy and various angular momenta is the first experimental evidence of the pairing reentrance in finite (hot rotating) nuclei.
In the study of GDR in highly excited nuclei, the PDM has been extended to include finite angular momentum. The results of calculations within the PDM are found in excellent agreement with the latest experimental data of GDR in the compound nucleus $^{88}$Mo.
Finally, an exact expression is derived to calculate the shear viscosity $\eta$ as a function of $T$ in finite nuclei directly from the GDR width and energy at zero and finite $T$. Based on this result, the values $\eta/s$ of specific shear viscosity in several medium and heavy nuclei were calculated and found to decrease with increasing $T$ to reach $(1.3 - 4)\times\hbar/(4\pi k_B)$ at $T =$ 5 MeV, that is almost the same value obtained for quark-gluon-plasma at $T >$ 170 MeV.
\end{abstract}

\section{Introduction}
Superfluid pairing plays an important role in the study of nuclear structure. In infinite systems the pairing gap $\Delta$ vanishes at a temperature $T_c\simeq 0.567\Delta(T=0)$, which is the critical temperature of the superfluid-normal (SN) phase transition. In deformed nuclei, the Coriolis 
force, which breaks the Cooper pairs, increases with the total angular momentum up to a certain critical angular momentum $J_c$, where all Cooper pairs are broken and the nucleus undergoes the SN phase transition, as suggested by the Mottelson-Valatin effect~\cite{Mottelson}. The combination of temperature and angular momentum effects causes the pairing reentrance phenomenon, which was predicted 50 years ago by Kammuri~\cite{Kammuri}. The physical interpretation of this phenomenon was given by Moretto~\cite{Moretto} as follows. In spherical nuclei at $T=$ 0, the total
angular momentum $J$ is made up by the nucleons from the broken
pairs, which occupy the single-particle levels around the Fermi surface and block them
against the scattered pairs. As $J$ increases, more levels around the Fermi surface are blocked, which make the pairing correlations decrease until
a sufficiently large total angular momentum $J_{c}$, where the pairing gap
$\Delta$ completely vanishes in infinite systems. The
increase of $T$ spreads the quasiparticles, which occupy the levels near the Fermi surface, farther away from it. Some levels become partially unoccupied, which are available for scattered pairs. 
As the result, when $T$ reaches a critical value $T_{1}$, the pairing correlations are energetically
favored, and the pairing gap reappears. As $T$ goes higher, the 
increase of a large number of quasiparticles quenches and eventually breaks down the
pairing gap at $T_{2}$ $(>T_{1})$.

It is now well known that, because of thermal fluctuations in finite nuclei, the paring gap does not vanish at $T\geq T_c$ and/or $J_c$, but moronically decreases with increasing $T$ and/or $J$. The pairing reentrance in finite nuclei is also modified so that the pairing gap remains finite after its first reappearance at finite $T$ and $J$. It has been shown within the phonon damping model (PDM)~\cite{PDM} that the presence of a thermal pairing gap at $T>T_c$ makes the width of the giant dipole resonance (GDR) remain nearly unchanged at low temperatures ($T\leq$ 1 MeV) in heavy nuclei. This lecture will show the most recent experimental facts that support this mechanism. It also demonstrates that the recently observed enhancement of the nuclear level densities~\cite{BARC,BARC4} is a clear evidence of pairing reentrance in $^{104}$Pd~\cite{Pd104}. In the description of the GDR width in hot rotating nuclei, the PDM has been extended to include the effect of angular momentum~\cite{PDMJ}. It will be shown that the predictions of the PDM agree well with the latest experimental data for the GDR in highly excited $^{88}$Mo. Finally, the GDR widths and energies predicted by the PDM and experimentally extracted are also used to calculate the shear viscosity of finite hot nuclei. 
\section{Pairing reentrance in hot rotating nuclei}
 The pairing pairing Hamiltonian for a spherical system rotating about the symmetry $z$ axis, where the total angular momentum $J$ is completely determined by its $z$-projection $M$ alone, has the form $H = \sum_k\epsilon_k(a_{+k}^{\dagger}a_{+k}+a_{-k}^{\dagger}a_{-k}) - G\sum_{kk'}{a_k^{\dagger}
a_{-k}^{\dagger}a_{-k'}a_{k'}}-\lambda\hat{N}-\omega\hat{M}$, where $a_{\pm k}^{\dagger}(a_{\pm k})$ are the creation (annihilation) operators of a particle (neutron or proton) in the $k$-th state, whereas $\epsilon_k$, $\lambda$ and $\omega$ are the single-particle energies, chemical potential and rotation frequency for an axially deformed system, respectively. This Hamiltonian was used to derive the FTBCS1 equations, which consist of a set of FTBCS (finite-temperature BCS) -based equations, corrected by the effects of quasiparticle-number fluctuations. The FTBCS1 equation for the pairing gap is written as a sum of the level-independent term $\Delta$ and level-dependent term $\delta\Delta_k$ as $\Delta_k = \Delta + \delta\Delta_k ~,$ where 
$\Delta=G\sum_{k'}{u_{k'}v_{k'}(1-n_{k'}^+-n_{k'}^-)}$, $\delta\Delta_k=G{\delta{\cal N}_k^2}u_kv_k/(1-n_k^+-n_k^-)$,
$u_k^2 = \left[1+(\epsilon_k-Gv_k^2-\lambda)/{E_k}\right]/2$, $v_k^2=\left[1-(\epsilon_k-Gv_k^2-\lambda)/{E_k}\right]/2$, 
$E_k = \sqrt{(\epsilon_k-Gv_k^2-\lambda)^2+\Delta_k^2}$, $n_k^{\pm} = [1+e^{(E_k \mp \omega m_k)/T}]^{-1}$.
with the quasiparticle-number fluctuations $\delta{\cal N}_k^2=(\delta{\cal N}_k^+)^2+(\delta{\cal N}_k^-)^2 = n_k^+(1-n_k^+)+n_k^-(1-n_k^-)$.
The equations for the particle number and total angular momentum are $N = 2\sum_k\left[v_k^2(1-n_k^+ -n_k^{-}) + \frac{1}{2}(n_k^\dagger+n_k^{-}) \right]$, $M = \sum_k m_k(n_k^+ - n_k^{-})$, respectively, with  $\pm m_k$ being the angular momentum projection quantum number.
By solving these FTBCS1 equations, one can easily calculate the total energy ${\cal E}$ and entropy $S$ of the system, from which the level density $\rho({\cal E},M)$ is calculated by using the invert Laplace transformation of the grand partition function~\cite{FTBCS1}.
The total level density $\rho({\cal E})$ of a system with the total energy ${\cal E}$ is a sum of $J$-dependent level densities (densities of states), namely $\rho({\cal E}) = \sum_{J}(2J+1)\rho({\cal E},J)$. The $J$-dependent level density  $\rho({\cal E},J)$ is calculated by differentiating $\rho({\cal E},M)$, that is $\rho({\cal E},J)=\rho({\cal E},M=J) - \rho({\cal E},M=J+1)~$.

Recently,  a series of measurements has been conducted at the Bhabha Atomic Research Center
for the reaction $^{12}$C + $^{93}$Nb $\rightarrow$ $^{105}$Ag$^{*}$ $\rightarrow$ $^{104}$Pd$^{*}$ $+$ p at the projectile energy $E(^{12}C)=$ 40 - 45 MeV~\cite{BARC,BARC4}. In these experiments the exclusive proton spectra in coincidence with a $\gamma$-ray multiplicity detector array reveal broad structures at higher multiplicities and energies. The obtained spectra including these broad structures can be fitted by using the statistical model only after multiplying the phenomenological level densities for $^{104}$Pd by an enhancement function, which depends on excitation energy and angular momentum. It has been suggested in Ref. ~\cite{BARC4} that this enhancement might come from the pairing reentrance.

\begin{figure}
    \center{
       \includegraphics[width=14cm]{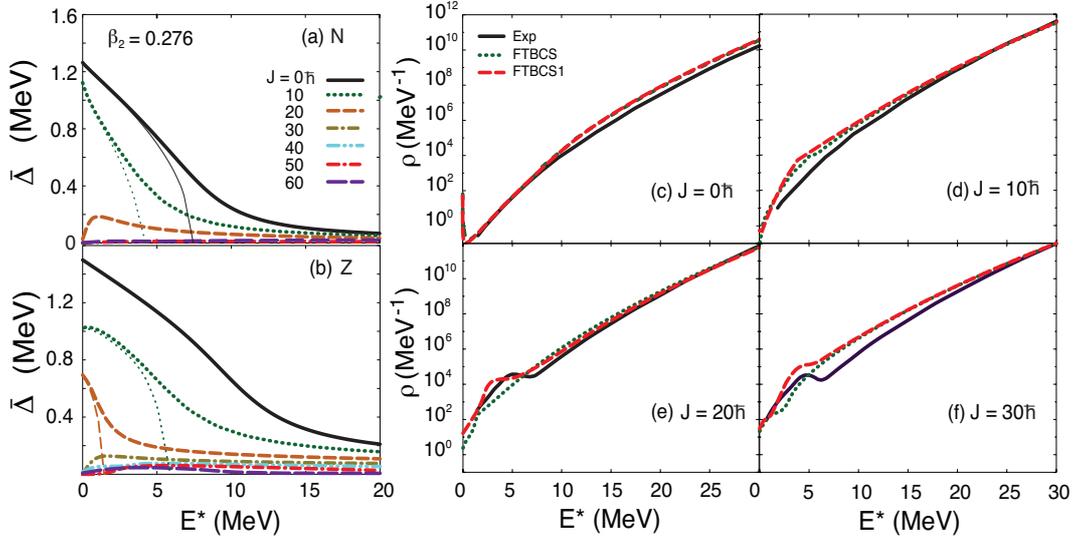}
       \caption{(Color online) Level-weighted pairing gaps $\overline{\Delta}$ for neutrons (a) and protons (b) as well as the level densities (c -- f) for $^{104}$Pd obtained within the FTBCS and FTBCS1 as functions of excitation energy $E^{*}$  at deformation parameter $\beta_2=$ 0.276 for different values of angular momentum $J$. In (a) and (b) the thin and thick lines denote the FTBCS and FTBCS1 pairing gaps, respectively. The parameters for pairing interaction are $G_N =$ 0.1737 MeV, $G_Z=$ 0.2152 MeV. In (c) -- (f), the dotted and dashed lines stand for the level densities $(2J+1)\rho({\cal E},J)$ predicted by the FTBCS and FTBCS1, respectively, whereas the solid lines are the empirical level densities. \label{prolate}}
       }
    \end{figure}

The level-weighted pairing gaps $\overline{\Delta}\equiv\sum_k\Delta_k/{\cal N}$ (with ${\cal N}$ being the sum of all levels) obtained within the conventional FTBCS show no pairing reentrance. The FTBCS neutron and proton gaps shown in panels (a) and (b) of Fig. \ref{prolate} decrease with increasing $E^{*}$ (or $T$) and collapses at some $E^{*}_c$, whose value decreases as $J$ increases. At the same time, the FTBCS1 gaps do not vanish at $E^{*}>$ 0. At $J<$ 20$\hbar$ they decrease monotonically as $E^{*}$ increases and remain finite even at $E^{*}>$ 15 MeV.  The pairing reentrance shows up at $J\geq$ 20$\hbar$ for neutrons and at $J\geq$ 30$\hbar$ for protons, where the gaps first increase with $E^{*}$ then decrease as $E^{*}$ increases further but do not vanish even at high $E^{*}$. Since this pairing reentrance takes place at rather high values of $J$, where the pairing gaps already vanish within the FTBCS, its origin resides in the quasiparticle number fluctuations $\delta{\cal N}_k^2$ in the FTBCS1 gap equation.

Our preliminary calculations in Ref.~\cite{Pd104} show that the corresponding level densities obtained within the FTBCS are smooth for all values of $E^{*}$ and $J$, whereas those predicted by the FTBCS1 have a local enhancement at 2 $< E^{*}<$ 4.5 MeV for all values of $J$ under consideration. For $\beta_2=$ 0.276, the FTBCS1 predictions agree fairly well with the empirical level densities obtained at $J\leq$ 30$\hbar$ [See Figs. \ref{prolate} (c) -- \ref{prolate} (f)]. For $\beta_2= -0.3$, a better agreement between theoretical predictions and empirical level densities is seen at $J>$ 30$\hbar$ (not shown). The fact that theoretical and empirical level densities agree well for $J\leq$ 30$\hbar$ at $\beta_2=$ 0.276, and for $J>$ 30$\hbar$ at $\beta_2= -0.3$ is a clear indication of a transition from a prolate shape at $J\leq$ 30$\hbar$ to an oblate shape at higher $J$. 

\section{Damping of GDR in highly excited nuclei}
\label{TJ}
The PDM Hamiltonian includes the 
independent single-particle (quasiparticle) field, GDR phonon field, and the coupling between 
them~\cite{PDM}.   The single-particle energies $\epsilon_k$ are obtained within the Woods-Saxon potentials. The GDR width $\Gamma(T)$ consists of the quantal width, $\Gamma_{\rm Q}$, and thermal width, $\Gamma_{\rm T}$, as $\Gamma(T)=\Gamma_{\rm Q}+\Gamma_{\rm T}$.
Including thermal pairing, the quantal and thermal widths 
are given as $\Gamma_{\rm Q}=2\gamma_Q(E_{GDR})=2\pi F_{1}^{2}\sum_{ph}[u_{ph}^{(+)}]^{2}(1-n_{p}-n_{h})
\delta[E_{\rm GDR}-E_{p}-E_{h}]~$,
and
$\Gamma_{\rm T}=2\gamma_T(E_{GDR})=2\pi F_{2}^{2}\sum_{s>s'}[v_{ss'}^{(-)}]^{2}(n_{s'}-n_{s})
\delta[E_{\rm GDR}-E_{s}+E_{s'}]~,$ 
where $u_{ph}^{(+)} = u_pv_h+u_hv_p$, $v_{ss'}^{(-)}=u_su_{s'}-v_sv_{s'}$ ($ss' = pp', hh'$), $n_k$ are quasiparticle occupations numbers, which are well approximated with the Fermi-Dirac distribution for
independent quasiparticles, $n_k = [\exp(E_k/T)+1]^{-1}$. The parameter $F_1$ is chosen so that $\Gamma_Q$ at $T=$ 0 is equal to GDR's width at $T=$ 0, whereas the parameter $F_2$ is chosen so that, with varying $T$,  the GDR energy $E_{GDR}$ does not change significantly. The latter is found as the solution of the equation $E_{GDR} - \omega_{q}-P_q(E_{GDR})=0$, where $\omega_q$ is the energy of the GDR phonon before the coupling between the phonon and single-particle mean fields is switched on, and $P_q(\omega)$ is the polarization operator owing to this coupling (Its expression is given in Ref. \cite{PDM}). The GDR strength function is found as $S_{q}(\omega) = (1/\pi)[\gamma_Q(\omega) + \gamma_T(\omega)]/\{(\omega-E_{GDR})^2+[\gamma_Q(\omega) + \gamma_T(\omega)]^{2}\}~.$
The representation $\delta(x) =\lim_{\varepsilon\rightarrow 0}\varepsilon/[\pi(x^{2}+\varepsilon^2)]$ is used for the $\delta$-function with $\epsilon=$ 0.5 MeV in numerical calculations.
\begin{figure}
\center{
\includegraphics[width=14cm]{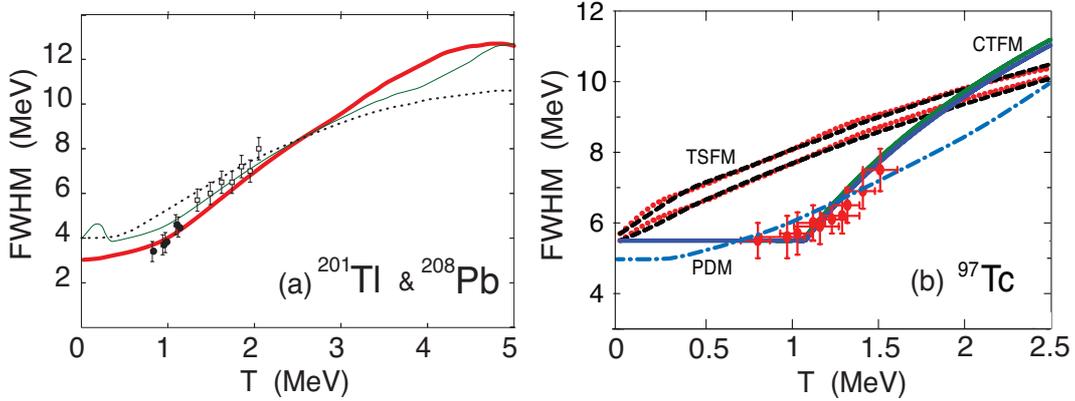}}
\caption{GDR width predicted within the PDM as a function of temperature for $^{201}$Tl, $^{208}$Pb (a) and $^{97}$Tc (b). In (a) the dotted and solid lines denotes the GDR width in $^{208}$Pb and $^{201}$Tl, respectively. The corresponding experimental data are shown as open boxes and full circles, respectively. In (b) the PDM prediction (dash-dotted line) is compared with the experimental data and predictions of the thermal shape fluctuation model (TSFM) with shell effect (dotted lines) and without shell effect (dashed lines) for J = 0$\hbar$ (lower) and J = 30$\hbar$ (upper) as well as those of the phenomenological model (continuous lines) for J = 10$\hbar$ (lower) and J = 20$\hbar$ (upper).
\label{width}}
\end{figure}
The GDR widths predicted by the PDM, thermal shape fluctuation model (TSFM), and the phenomenological model called critical temperature included fluctuation model (CTFM) are shown in Fig. \ref{width}  in comparison with the experimental systematics for $^{97}$Tc~\cite{Tc97}, $^{201}$Tl~\cite{Tl201}, and $^{208}$Pb~\cite{Schiller}. The PDM results for open shell nuclei include the effect of non-vanishing thermal pairing gap owing to thermal fluctuations in finite nuclei. The PDM is the only up-to-date semi-microscopic model that is able to describe well the experimental data in the entire temperature region including $T\leq$ 1 MeV as well as high $T$, where the GDR width seems to saturate. 

\begin{figure}
\center{
\includegraphics[width=10cm]{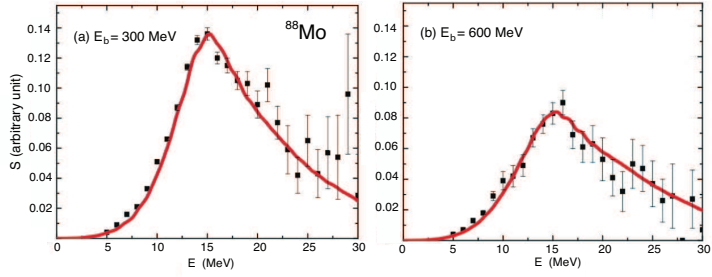}}
\caption{PDM predictions for the GDR line shapes in $^{88}$Mo (solid lines) in comparison with the experimental data at the beam energies of 300 MeV (a) and 600 MeV (b), which correspond to the average temperatures of 2.04 and 3.06 MeV, respectively. The average angular momentum is equal to 38$\hbar$ (Adapted from Figs. 5.5 and 5.6 of Ref. \cite{Ciemala}).
\label{Mo88}}
\end{figure}

To describe the non-collective rotation of a spherical nucleus, the $z$-projection $M$ of the total angular momentum $J$ is added into the PDM Hamiltonian as  $- \gamma\hat{M}$, where $\gamma$ is the rotation frequency~\cite{PDMJ}. The latter and the chemical potential are defined, in the absence of pairing,  from the equation $M = \sum_k m_k(f_{k}^{+}-f_{k}^{-})~,$ and  $N =\sum_k(f_{k}^{+}+f_{k}^{-})~$, where $N$ is the particle number and $f_k^{\pm}$ are the single-particle occupation numbers, $f_{k}^{\pm} =1/[\exp(\beta E_k^{\mp})+1]$, and $E_k^{\mp} = \epsilon_k-\lambda\mp\gamma m_k~$. The prediction of the PDM for the GDR line shape agrees remarkably well with the latest experimental date for $^{88}$Mo at high $T$ and $J$~\cite{Ciemala}, as shown in Fig. \ref{Mo88}.
\section{Shear viscosity of hot nuclei}
\label{visco}
For the application of hydrodynamics to 
nuclear system, the quantum mechanical 
uncertainty principle requires a finite viscosity for any thermal 
fluid. Kovtun, Son and Starinets (KSS)~\cite{KSS} conjectured
that the specific viscosity, which is the ratio $\eta/s$ of shear viscosity $\eta$ to the entropy 
volume density $s$, is bounded at the lower end for all fluids by the universal value, which is called the KSS bound (or unit), 
${\eta}/{s}= {\hbar}/(4\pi k_{B})$.

According to collective theories, one of the fundamental 
explanations for the giant resonance damping is the friction term (viscosity) 
of the neutron and proton fluids.  By using the Green-Kubo's relation, Ref. \cite{visco} expresses the shear viscosity $\eta(T)$ 
at finite $T$ in terms of the GDR's energies and widths at zero and finite $T$ as
\begin{equation}
\eta(T)=\eta(0)\frac{\Gamma(T)}{\Gamma(0)}
\frac{E_{GDR}(0)^{2}+[\Gamma(0)/2]^{2}}{E_{GDR}(T)^{2}+[\Gamma(T)/2]^{2}}~.
\label{eta1}
\end{equation}
\begin{figure}
     \center{\includegraphics[width=16.0cm]{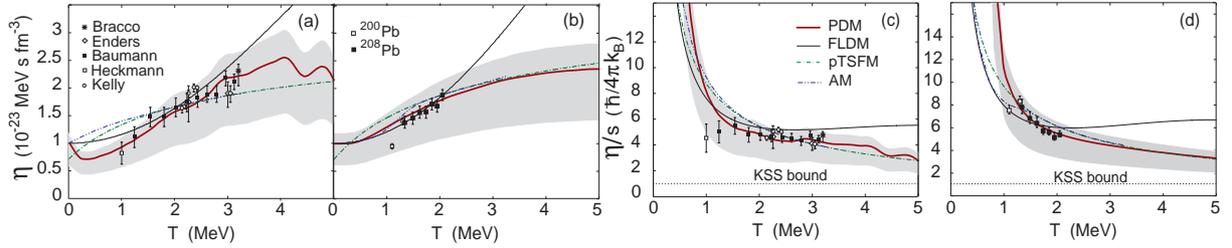}
     \caption{Shear viscosity $\eta(T)$ [(a) and (b)] and ratio 
     $\eta/s$ [(c) and (d)] as functions of $T$ for nuclei in tin [(a) and (c)], and lead [(b) and (d)] regions. The  
     gray areas are the PDM predictions by using $0.6u\leq\eta(0)\leq 
     1.2u$ with $u=$ 10$^{-23}$ Mev s fm$^{-3}$.}
     \label{eta&ratio}}
\end{figure}

The predictions for the shear viscosity $\eta$ and the specific viscosity $\eta/s$ by several theoretical models 
for $^{120}$Sn and $^{208}$Pb are plotted as functions of $T$ 
in Fig. \ref{eta&ratio} in comparison with the empirical results. 
The latter are extracted from the 
experimental systematics for GDR in tin and lead regions~
\cite{Schiller} making use of Eq. (\ref{eta1}).  It is seen from Fig. \ref{eta&ratio} that the predictions by the PDM agree best 
with the empirical results. The ratio $\eta/s$ decreases sharply with 
increasing $T$ up to $T\sim$ 1.5 MeV, starting from which the decrease 
gradually slows down to reach (2 - 3) KSS units 
at $T=$ 5 MeV. The fermi-liquid drop model (FLDM) has a similar trend up to 
$T\sim$ 2 - 3 MeV, but at higher $T$ ($T>$ 3 MeV for $^{120}$Sn or 2 MeV for 
$^{208}$Pb) both $\eta$ and $\eta/s$ with $T$ sharply increase. 
At $T=$ 5 MeV the FLDM model predicts the ratio $\eta/s$ within (3.7 - 6.5) KSS units, that is 
roughly 1.5 -- 2 times larger than the PDM predictions. The TSFM in two versions as adiabatic model (AM) and phenomenological TSFM (pTSFM) 
show a similar trend for $\eta$ and $\eta/s$, provided $\eta(0)$ in the pTSFM 
calculations is reduced to 0.72$u$. However both versions of TSFM 
overestimate $\eta$ at $T<$ 1.5 MeV. Based on these results and on a model-independent estimation, it is concluded that
the PDM predicts 1.3 $\leq\eta/s\leq$ 4 KSS units for medium and heavy nuclei at $T=$ 5 MeV. This is about (3 - 5) times smaller 
(and of much less uncertainty) than the value between (4 - 19) KSS units predicted by 
the FLDM for heavy nuclei~\cite{Auerbach}, where the same lower value $\eta(0)=$0.6$u$ was used.
\section{Conclusions}
This lecture has confirmed that thermal pairing included in the phonon damping model (PDM) is indeed responsible for the nearly constant width of the giant dipole resonance (GDR) at low temperature $T <$ 1 MeV. It also demonstrates that the enhancement, which has been observed in the recent experimentally extracted nuclear level densities in $^{104}$Pd at low excitation energy and various angular momenta, is the first experimental evidence of the pairing reentrance in finite (hot rotating) nuclei. The extension of the PDM to finite angular momentum offers predictions in excellent agreement with the latest experimental data of GDR in the compound nucleus $^{88}$Mo. The specific shear viscosity $\eta/s$ in several medium and heavy nuclei is calculated from the GDR widths and energies at zero and finite temperatures and found to decrease with increasing $T$ to reach $(1.3 - 4)\times\hbar/(4\pi k_B)$ at $T =$ 5 MeV, which is almost the same value obtained for quark-gluon-plasma at $T >$ 170 MeV.

The calculations were carried out on the RIKEN Integrated Cluster of Clusters (RICC) system. Thanks are due to M. Ciemala for permission to use the experimental GDR line shapes in $^{88}$Mo~\cite{Ciemala}.

\end{document}